\documentclass[12pt,manuscript=article]{achemso}
\setkeys{acs}{maxauthors = 0}
\usepackage{graphics,epsfig,color,amsmath,appendix}
\usepackage{subcaption}
\usepackage{comment}
\usepackage{setspace}
\usepackage{hyperref}
\usepackage{siunitx}
\newcommand{\bm}[1]{{\mathbf{#1}}}

\title{Relative Binding Free Energy Estimation of Congeneric Ligands and Macromolecular Mutants with the Alchemical Transfer with Coordinate Swapping Method}

\author{Emilio Gallicchio}
\email{egallicchio@brooklyn.cuny.edu}
\affiliation{Department of Chemistry and Biochemistry, Brooklyn College of the City University of New York, New York, NY, 11210}
\alsoaffiliation{Ph.D. Program in Chemistry, The Graduate Center of the City University of New York, New York, NY, 10016}
\alsoaffiliation{Ph.D. Program in Biochemistry, The Graduate Center of the City University of New York, New York, NY, 10016}
\date{}

\begin{document}

\maketitle
\begin{abstract}
  We present the Alchemical Transfer with Coordinate Swapping (ATS) method to enable the calculation of the relative binding free energies between large congeneric ligands and single-point mutant peptides to protein receptors with the Alchemical Transfer Method (ATM) framework. Similarly to ATM, the new method implements the alchemical transformation as a coordinate transformation, and works with any unmodified force fields and standard chemical topologies. Unlike ATM, which transfers the whole ligands in and out of the receptor binding site, ATS limits the magnitude of the alchemical perturbation by transferring only the portion of the molecules that differ between the the bound and unbound ligands. The common region of the two ligands, which can be arbitrarily large, is unchanged and does not contribute to the magnitude and statistical fluctuations of the perturbation energy. Internally, the coordinates of the atoms of the common regions are swapped to maintain the integrity of the covalent bonding data structures of the molecular dynamics engine. The work successfully validates the method on protein-ligand and protein-peptide RBFE benchmarks. This advance paves the road for the application of the relative binding free energy Alchemical Transfer Method protocol to study the effect of protein and nucleic acid mutations on the binding affinity and specificity of macromolecular complexes. 
\end{abstract}

\section{Introduction}

Increasingly, lead optimization draws data from molecular computer simulations to speed up the drug development process. Alchemical calculations are used to estimate the Relative Binding Free Energies (RBFE) of protein-ligand complexes to rank their binding affinities and prioritize their evaluation.\cite{GallicchioSAMPL4,wang2015accurate,zou2019blinded,schindler2020large,lee2020alchemical,kuhn2020assessment,gapsys2020large,bieniek2021ties,hahn2022bestpractices,gapsys2022pre,sabanes2023validation,cournia2017relative}.   RBFE alchemical models estimate the difference in standard binding free energies, $\Delta\Delta G^\circ_b$, between two ligands binding to the same receptor by considering non-physical processes that progressively modify the system's potential energy function in such a way that at the beginning it describes the receptor bound to the first ligand and at the end it describes the receptor bound to the second.\cite{Jorgensen2004,cournia2017relative,Mey2020Best,azimi2022relative} The alchemical process typically involves several intermediate thermodynamic steps. In popular double-decoupling RBFE implementations,\cite{Gilson:Given:Bush:McCammon:97,wang2015accurate,lee2020alchemical} for example, the RBFE is calculated as the difference of the relative free energies of coupling the ligands to the solution and receptor environments from a gas-phase reference state.  

Alchemical RBFE implementations can be broadly classified by their ligands' representations. Methods that we will refer to as \emph{dual-topology} employ individual standard chemical topologies for the two ligands, and the alchemical transformations consist of decoupling and coupling each ligand are a whole to their original and target environments--vacuum, the solution, or the receptor depending on the specific step in the thermodynamic cycle.\cite{Michel2007,riniker2011comparison} When comparing congeneric ligands that differ by a small peripheral group of a larger common core (R-group transformations), \emph{single-topology} RBFE approaches can be employed that hold one alchemical topology with a shared common core attached to the R-groups designed to map the topology of one ligand to the other as the alchemical transformation progresses.\cite{pearlman1994comparison,fleck2021dummy} Because they track only the potential energy perturbation due to the change of the R-group rather than the whole ligand, single-topology implementations tend to be more efficient than dual-topology implementations, especially when the variable R-group is much smaller than the common core. 

It should be noted that the terms single- and dual-topology have sometimes been used to distinguish alchemical topology implementations depending on the use of dummy atoms\cite{fleck2021dummy} to describe atoms present in one ligand and absent in the other.\cite{shirts2013introduction,Mey2020Best} Here, instead, we use the term dual-topology to specifically refer to methods where the two ligands are described by distinct standard chemical topologies whose structure and composition are not affected by the alchemical transformation.\cite{Michel2007,riniker2011comparison,rocklin2013separated} However they are called, the distinction between the two methods is that single-topology methods limit the alchemical perturbation to the variable region of the ligand pair, whereas in the dual-topology approach, the alchemical transformation affects the intermolecular interactions of all of the atoms in the ligands whether they differ between the two ligands or not. The result is that the magnitude of the dual-topology alchemical transformation scales as the size of the ligands rather than, as in single-topology approaches, the magnitude of the difference between them. As a result, dual-topology workflows generally display lower computational efficiency than single-topology for treating congeneric ligand libraries differing in small R-group modifications and are not applicable to study modifications of macromolecular ligands such as proteins and peptides.

Despite their lower performance and inability to treat macromolecular ligands, dual-topology alchemical approaches have appealing features. Because they utilize standard chemical topologies, dual-topology implementations require fewer modifications of molecular dynamics engines, can support a wider range of force fields,\cite{sabanes2024enhancing} and are easier to maintain than single-topology modules.\cite{zou2019blinded} Furthermore,  because they do not restrict the nature of the ligands and do not require atom-mapping,\cite{LOMAP} dual-topology approaches can more easily tackle a wider range of ligand variations, such as scaffold-hopping transformations,\cite{wang2017accurate,vilseck2019overcoming,zou2021scaffold} with a high level of automation.\cite{azimi2022relative,chen2024performance} 

We have recently developed the Alchemical Transfer Method (ATM) for relative binding free energy estimation (ATM-RBFE),\cite{azimi2022relative} which implements the alchemical transformation as a coordinate transformation that switches the positions of the bound and unbound ligands within a dual-topology framework. The software implementation of ATM with the OpenMM molecular dynamics engine is freely available\cite{eastman2023openmm,AToM-OpenMM-24}. We and others have deployed it in large-scale RBFE campaigns with standard force fields\cite{chen2024performance,sabanes2023validation} and advanced neural network potentials.\cite{sabanes2024enhancing} We have extended the method to incorporate accelerated conformational sampling protocols,\cite{khuttan2024make} to treat multiple binding poses,\cite{khuttan2023taming} and to model binding specificity.\cite{azimi2024binding} However, because it is based on a dual-topology approach, ATM-RBFE cannot access the higher computational efficiency of single-topology approaches for R-group transformations and is not currently applicable to large ligands and macromolecular mutants.

In this work, we present an extension of the ATM-RBFE protocol we call Alchemical Transfer with coordinate Swapping (ATS-RBFE), which applies to any R-group transformation where the single-decoupling approach applies. The alchemical perturbation is implemented similarly to the ATM-RBFE method, except that it translates only the positions of the variable R-groups of the bound and unbound ligands rather than the whole ligands. In addition, the coordinates of the corresponding atoms of the common core of the two ligands are swapped to preserve the integrity of the representation of the chemical topologies and minimize changes in covalent interactions. The method retains the simplicity and favorable features of ATM-RBFE's dual-topology strategy while affording the greater computational efficiency and applicability to macromolecular ligands of the single-topology approach. 

We first present the method from a statistical mechanics perspective and prove in the Appendix that is mathematically exact. We then describe the software implementation and its application to protein-ligand and protein-peptide RBFE benchmarks. All of the results confirm the correctness of the novel ATS-RBFE method and its applicability to estimating RBFEs for R-group transformations between large ligands and single-point mutants.  

\section{Theory and Methods}

\subsection{The Alchemical Transfer Method for Relative Binding Free Energy Estimation}

We briefly summarize the Alchemical Transfer Method (ATM) for relative binding free energy estimation (RBFE), which is the basis for the Alchemical Transfer with coordinate Swapping (ATS) method presented here. A thorough account of ATM-RBFE and its applications is available in published works.\cite{azimi2022relative,sabanes2023validation,chen2024performance,khuttan2024make} 

Consider a pair of compounds, $A$ and $B$, binding to the same receptor $R$. The system is prepared so that the first ligand is bound to the receptor, and the second is placed in the solvent bulk and displaced by the first by some displacement vector $d$. The ligands are kept near these positions by suitable flat-bottom restraining potentials that describe the chosen extent of the receptor binding site.\cite{azimi2022relative} The standard binding free energy $\Delta \Delta G_b^\circ = \Delta G_b^\circ(B) - \Delta G_b^\circ(A)$ of the second ligand relative to the first is then expressed as
\begin{equation}
    \Delta \Delta G_b^\circ = - (1/\beta) \ln \langle e^{-\beta u(\bm{r})}  \rangle_{RA + B}
    \label{eq:G-rbfe}
\end{equation}
where $\beta = 1/(k_B T)$ is the inverse temperature, $k_B$ is Boltzmann's constant, $\bm{r}$ represents the coordinates of the system's atoms, $u$ is the ATM perturbation energy defined below, and the averaging $\langle \ldots \rangle$ is performed in the ensemble in which ligand $A$ is bound to the receptor (the state $RA + B$). The ATM perturbation energy
\begin{equation}
    u(\bm{r}) = U( x_{R}, \bm{r}_{A} + \bm{d},  \bm{r}_{B}-\bm{d},  \bm{r}_{S}) - U(x_{R}, \bm{r}_{A},  \bm{r}_{B},  \bm{r}_{S}) 
    \label{eq:u-def}
\end{equation}
where $\bm{r} = (x_{R}, \bm{r}_{A},  \bm{r}_{B},  \bm{r}_{S})$ with $x_R$, $\bm{r}_A$, $\bm{r}_B$, and $\bm{r}_S$ being the internal coordinates of the receptor, the first ligand, the second ligand, and of the solvent atoms, respectively, is defined as the change of the system's potential energy $U$ resulting from translating the coordinates of the atoms of $A$ by the fixed displacement vector $\bm{d}$ and simultaneously displacing the coordinates of the second ligand in the opposite direction. This coordinate transformation results in the second ligand being bound to the receptor and the first ligand, initially bound to the receptor, being into the solvent bulk.

The ensemble average in Eq.\ (\ref{eq:G-rbfe}) is calculated by standard $\lambda$-parameterized stratification strategy, multi-state free energy estimation, and the interpolating alchemical potential energy function
\begin{equation}
    U_\lambda(\bm{r}) = U(\bm{r}) + W_\lambda[u(\bm{r})]
    \label{eq:pot-func}
\end{equation}
where the soft-core softplus alchemical perturbation function is  
\begin{equation}
  W_{\lambda}[u]=\frac{\lambda_{2}-\lambda_{1}}{\alpha}\ln\left\{1+e^{-\alpha [u_{\rm sc}(u)-u_{0}]}\right\}+\lambda_{2}u_{\rm sc}(u) .\, ,
  \label{eq:softplus-function}
\end{equation}
where the parameters $\lambda_{2}$, $\lambda_{1}$, $\alpha$, $u_{0}$ are functions of $\lambda$ (see Computational Details),
and the function
\begin{equation}
  u_{\rm sc}(u)=
\begin{cases}
u & u \le u_c \\
(u_{\rm max} - u_c ) f_{\rm sc}\left[\frac{u-u_c}{u_{\rm max}-u_c}\right] + u_c & u > u_c
\end{cases}
\label{eq:soft-core-general}
\end{equation}
with
\begin{equation}
f_{\rm sc}(y) = \frac{z(y)^{a}-1}{z(y)^{a}+1} \label{eq:rat-sc} \, ,
\end{equation}
and
\begin{equation}
    z(y)=1+2 y/a + 2 (y/a)^2
\end{equation}
is the soft-core perturbation energy function designed to avoid singularities near the initial state of the alchemical transformation.\cite{pal2019perturbation,khuttan2021alchemical} The soft-core parameters $u_{\rm max}$, $u_c$, and $a$ used in this work are given in Computational Details.

\subsection{The Alchemical Transfer with Coordinate Swapping Protocol for Relative Binding Free Energy Estimation}

As described above, in the original ATM method, all of the coordinates of the atoms of the ligands, including those in common substructures, are perturbed by the displacement transformation. However, alchemical relative binding free energy calculations often involve congeneric ligands that share large portions of their structures. Here, we outline an alchemical transfer free energy protocol optimized for these situations based on the displacement of the variable portion of the two ligands and the swapping of coordinates of the atoms in the common substructure.

\begin{figure}
    \centering
    \includegraphics[scale=0.25]{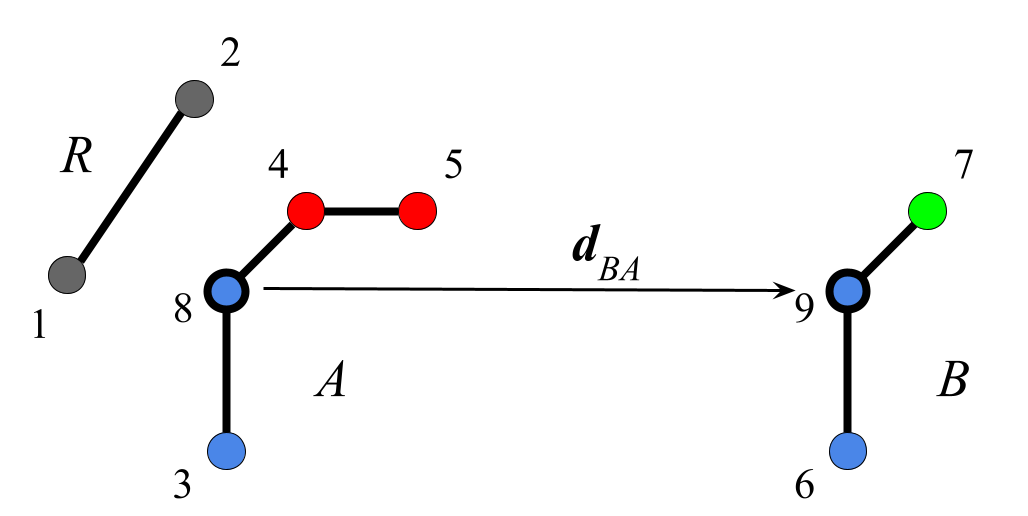}
    \caption{\label{fig:two-ligs} Schematic diagram illustrating the ATS coordinate transformation for the estimation of the relative binding free energy between atoms $A$ and $B$. The black atoms represent the receptor $R$; the blue atoms represent the common region of the ligand pair, and the red and green atoms represent the variable peripheral R-group of the two ligands. The $\bm{d}_{BA}$ vector is the displacement between the anchor atoms denoted by a heavy line border. The coordinate transformation involves translating the red atoms of ligand $A$ by $\bm{d}_{BA}$ and the green atom of ligand $B$ by the opposite displacement and simultaneously swapping the coordinates of the corresponding atoms of the common region (the blue atoms). }
\end{figure}

Consider, for example, the two ligands $A$ and $B$ schematically drawn in Figure \ref{fig:two-ligs} with their common regions, $A'$ and $B'$, denoted by the blue atoms and the variable regions, $A''$ and $B''$, denoted by the red and green atoms. The gray atoms denote schematically the receptor (R). Atoms 8 and 9, denoted by the heavier borderline (the indexing of the atoms is explained in the Appendix),  are the anchoring atoms of the variable region to the common region. We also assume that there is a one-to-one mapping between the atoms of the common regions of the two ligands. For the case of Figure \ref{fig:two-ligs}, for example, atom 3 of ligand $A$ is mapped to atom $6$ of $B$, and atom $8$ is mapped to atom $9$, and vice versa.

As shown in the Appendix, the ratio of binding constants (functionally equivalent to the relative binding free energy, RBFE) can be expressed by the ensemble average 
\begin{equation}
      \frac{K_b(B)}{K_b(A)} = \langle e^{-\beta u} \rangle_{RA+B}
      \label{eq:Kbrel-average}
\end{equation}
where the averaging is performed over the ensemble $RA+B$, where $A$ is bound to $R$, and the perturbation energy $u$ is the change in potential energy for swapping the coordinates of the atoms in the common regions of the two ligands, and simultaneously displacing the coordinates of the atoms of the variable region of $A$ to the position of the anchoring atom of $B$, and those of $B$ to the position of the anchoring atom of $A$. Specifically, the perturbation energy is written as 
\begin{equation}
u = U_{RB+A} - U_{RA+B}
\label{eq:pert-func-rel}
\end{equation}
where $U$ is the potential energy function of the system,
\begin{equation}
    U_{RA+B} = U(x_{R}, \bm{r}_{A'},  \bm{r}_{A''},  \bm{r}_{B'}, \bm{r}_{B''}, \bm{r}_S)
    \label{eq:pot-Abound}
\end{equation}
where $x_R$ represents the internal degrees of freedom of the receptor, $\bm{r}_{A'}$ and $\bm{r}_{A''}$ are the coordinates of the atoms of the common and variable regions of ligand $A$, respectively, and similarly for $B$, and $\bm{r}_S$ are the coordinates of the solvent molecules, is the potential energy of the system when ligand $A$ is bound to the receptor and ligand $B$ is in the solvent bulk, and
\begin{equation}
U_{RB+A} = U(x_{R}, \bm{r}_{B'},  \bm{r}_{A''}+\bm{d_{BA}},  \bm{r}_{A'}, \bm{r}_{B''}-\bm{d}_{BA}, \bm{r}_S)
\label{eq:pot-Bbound}
\end{equation}
is the potential energy when $B$ is bound to the receptor. The latter is obtained from the $RA+B$ state by swapping the coordinates $\bm{r}_{A'}$ and $\bm{r}_{B'}$ of the corresponding atoms of the common region and rigidly translating the coordinates of the variable atoms of $A$ by the displacement vector, $\bm{d}_{BA}$, of the position of the anchoring atom of $B$ relative to the anchoring atom of $A$, and applying the opposite displacement to the coordinates of the variable atoms of $B$. The ensemble average $\langle \ldots \rangle_{RA+B}$ is carried out in the $RA+B$ state, that is, while ligand $A$ is the binding site and ligand $B$ is in the bulk.

Essentially, this protocol, which we name Alchemical Transfer with coordinate Swapping (ATS), corresponds to the Alchemical Transfer RBFE protocol (see above) when only the variable regions of the ligands are transferred in and out of the binding site. Because the atoms of the common region of one ligand are mapped to equivalent atoms of the other ligand, the atoms of the common region effectively remain in place due to the coordinate swapping. The protocol is amenable to molecular dynamics sampling because the covalent interactions (bonds, angles, and torsions) between the variable and common regions are approximately preserved when the atoms are simultaneously displaced and swapped. For example, the bond between atoms $4$ and $8$ in Figure \ref{fig:two-ligs} remains the same when atom 8 takes the position of atom $9$, which is its mapped atom, and atom $4$ is translated by the displacement vector $\bm{d}_{BA}$ of atom $9$ relative to atom $8$, which are the two anchoring atoms in this case. The bonding relationship between the peripheral atoms of the variable region relative to the common region (such as atom $7$ in Figure \ref{fig:two-ligs}) is not necessarily unchanged due to the coordinate displacement and swapping transformation. However, it is expected that the change in bonding energy is small enough to introduce small statistical fluctuations and that the free energy estimator of Eq.\ (\ref{eq:Kbrel-average}) with the coordinate displacement and swapping transformation can yield converged free energy estimates in most cases of interest.

\subsection{Molecular Systems}

\begin{figure}
    \centering
    \includegraphics[scale=0.45]{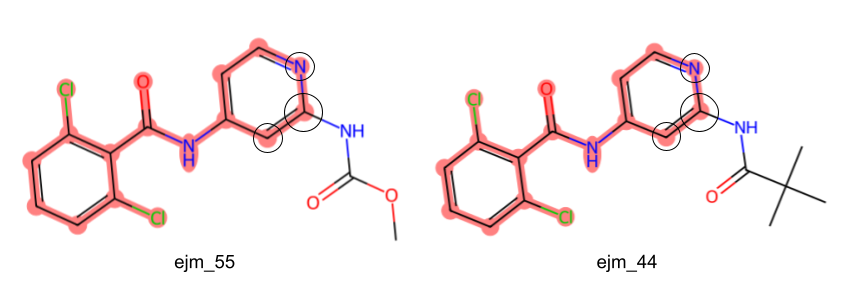}
    \caption{\label{fig:tyk2-ejm-55-44} A representative ligand pair from the TYK2 RBFE benchmark set.\cite{ross2023maximal} The common region is highlighted in red. The anchor atom is indicated by a large circle. The smaller circles indicate the two atoms that define the alignment frame together with the anchor atom. }
\end{figure}

In this work, we considered two molecular systems. The first is the TYK2 community benchmark set of 24 ligand pairs binding the TYK2 tyrosine protein kinase\cite{liang2013lead-1,liang2013lead-2} assembled by Schr\"{o}dinger.\cite{wang2017accurate,ross2023maximal,Schrodinger-benchmarks} This set is particularly suitable for testing the ATS-RBFE protocol because the ligands share a common core and differ only by a peripheral side-chain (Fig.~\ref{fig:tyk2-ejm-55-44}). Yet, their relatively small size allows a comparison with the standard ATM dual-topology RBFE protocol (see Results).

\begin{figure}
    \centering
    \includegraphics[scale=0.45]{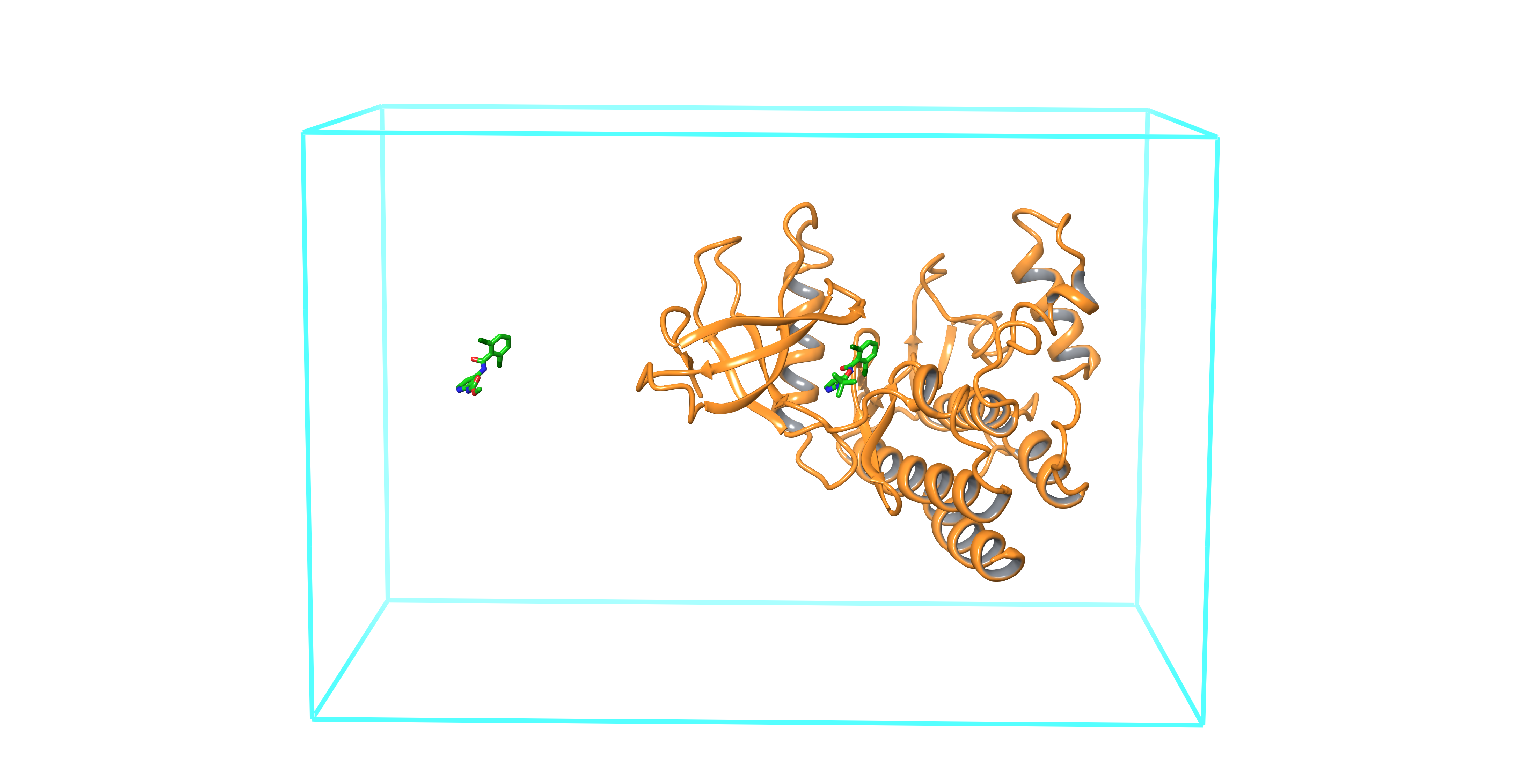}
    \caption{\label{fig:tyk2-setup} Illustration of the ATM and ATS simulation setups for the protein-ligand RBFE calculations. The simulation box is in cyan. The TYK2 protein is shown in orange ribbon. The ejm\_44 ligand (green) is bound to the receptor and the ejm\_55 ligand is in the solvent. The solvent molecules are not shown for clarity. In the ATM protocol, the alchemical perturbation involves translating the two ligands by a fixed displacement so that their positions are inverted. In the ATS protocol, the variable R-group is translated by the displacement between the two anchor atoms and simultaneously swapping the coordinates of the atoms of the common region (Fig.\ \ref{fig:tyk2-ejm-55-44}).}
\end{figure}

\begin{figure}
    \centering
    \includegraphics[scale=0.45]{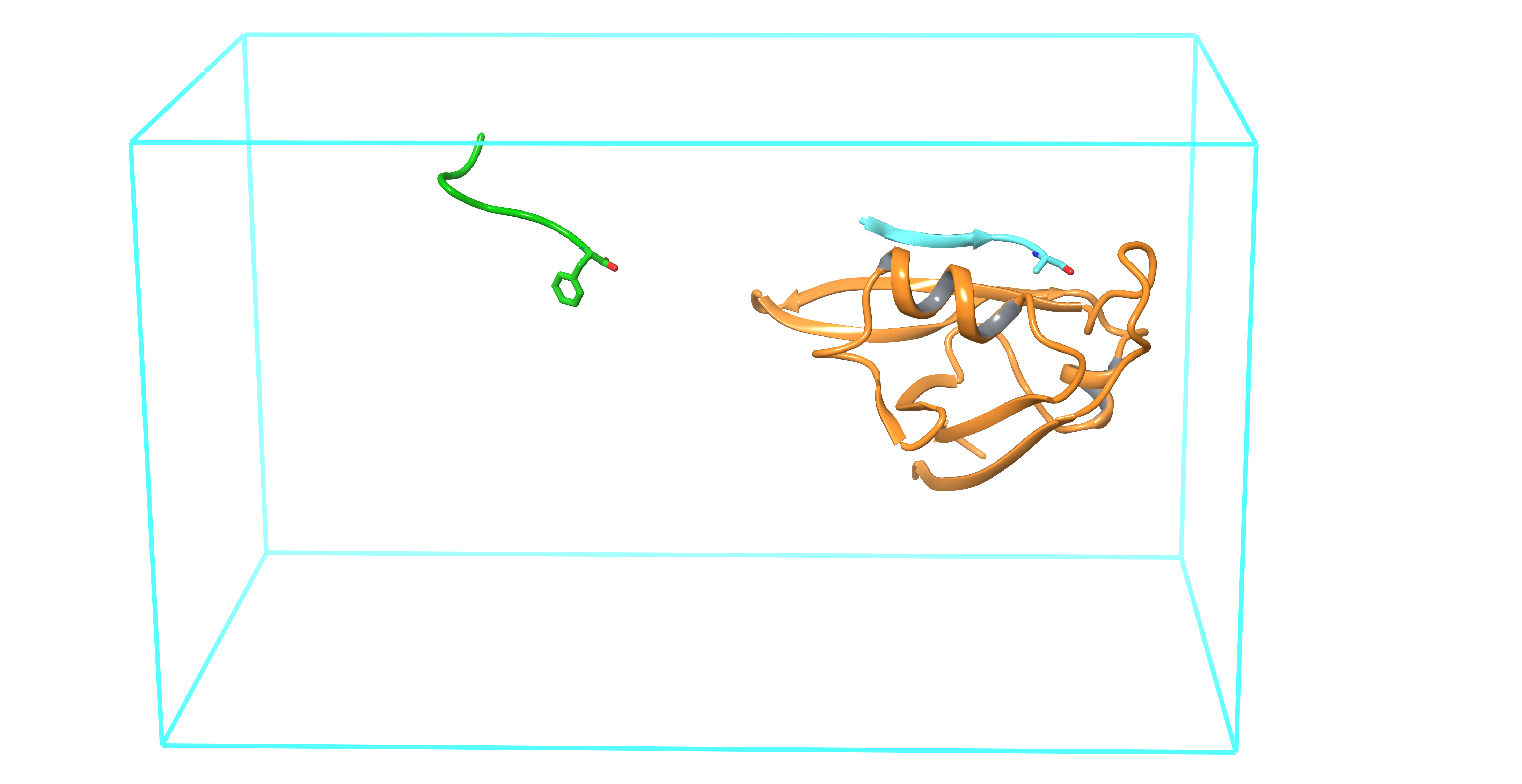}
    \caption{\label{fig:tiam-1-atmsetup} Illustration of the ATS simulation setup for the protein-peptide RBFE calculations. The simulation box is in cyan. The TIAM-1 PDZ domain receptor is shown in orange ribbon. The wild-type Syndecan-1 peptide (cyan) is bound to the receptor and the A0F mutant (green) is in the solvent. The solvent molecules are not shown for clarity. The mutated residues (alanine for the wild-type and phenylalanine for the mutant) are displayed in licorice representation. The alchemical perturbation involves translating the sidechain of each mutated residue by the displacement between the two C-$\alpha$ atoms and swapping the coordinates of the other atoms of the two peptides.}
\end{figure}

The second set includes the complexes between the TIAM-1 PDZ domain and four peptides differing by a single-point aminoacid mutation.\cite{shepherd2011distinct,liu2013structure,panel2018accurate,gallicchio2021comppeptsci} The wild-type peptide derived from the Syndecan-1 membrane receptor has sequence TKQEEFYA, and the mutants replace the C-terminal alanine with phenylalanine (F), methionine (M), and valine (V). The peptide forms a series of interactions with one of the $\beta$-sheet and a $\alpha$-helices of the PDZ domain. The mutated aminoacid occupies a recognition pocket that accommodates mostly hydrophobic sidechains (Figure \ref{fig:tiam-1-atmsetup}).  

\subsection{Simulation Settings}

We employed the structures of the TYK2 enzyme and the corresponding ligands from Schr\"{o}dinger's repository.\cite{ross2023maximal,Schrodinger-benchmarks} For the ATM-RBFE calculations, we employed the AToM-OpenMM setup workflow similar the CDK2 example in the AToM-OpenMM distribution\cite{AToM-OpenMM-24} with a ligand displacement of $45$ \AA\ in the negative $z$ direction, resulting in systems with the first ligand bound to the receptor and the second placed in the solvent (Figure \ref{fig:tyk2-ejm-55-44}). The Amber ff14SB force field\cite{maier2015ff14sb} was used for the protein and OpenFF 2.0\cite{boothroyd2023development} for the ligands. Quadratic flat-bottom positional restraints with a tolerance of $3.0$ \AA\ and a force constant of $25$ kcal/mol/\AA$^2$ were applied to the C-$\alpha$ atoms of TYK2. 

As in previous work,\cite{azimi2022relative,khuttan2023taming} we implemented the indicator function\cite{Gilson:Given:Bush:McCammon:97,Gallicchio2011adv} of the TYK2 complexes using a quadratic flat-bottom potential of tolerance $5$ \AA\ and force constant $25$ kcal/mol/\AA$^2$ between the centers of the receptor binding site and of the ligands. The second ligand was displaced to the binding site before applying the binding site restraint. The center of the receptor binding site region was set as the centroid of the C-$\alpha$ atoms of TYK2's residues 900--913,  926--930,   947, 950,  960--963, 976--989,  1023--1032, and 1038--1043. The center of the ligand was taken as the position of the anchor atom (see below). The ligand atoms of the common region indicated in Figure \ref{fig:two-ligs} were used as RBFE alignment atoms.\cite{azimi2022relative} The anchor atom labeled by a large circle was taken as the origin of the alignment frame. We employed an identical setup for the ATS-RBFE calculations, using the core highlighted in red in Figure \ref{fig:two-ligs} and using the atom labeled by a large circle as the anchor atom.

The protein-peptide systems were built similarly to the protein-ligand systems. TIAM-1 PDZ complexed with the wild-type Syndecan-1 peptide was prepared from the crystal structure (PDB is 4GVD)\cite{liu2013structure} using the protein preparation workflow in Maestro 2023-4 (Schr\"{o}dinger, LLC). The mutated peptides were created from the wild-type using the Mutate Residue facility in Maestro. The pose of the mutated residue and the adjacent residue (positions 0 and 1) were energy-minimized in the receptor binding site to avoid atomic clashes. The unbound peptide was placed in the solvent by displacing it by 40 \AA\ in the $y$-direction  from the bound pose (Figure \ref{fig:tiam-1-atmsetup}). The system was solvated in TIP3P water and neutralizing ions with a $10$ \AA\ padding in each direction using the Modeller facility of OpenMM.\cite{eastman2023openmm} 

The ATS-RBFE protocol was applied to the protein-peptide systems, taking the sidechain of the mutated residues as the variable region and the rest of the peptide as the common region. The C-$\alpha$ atom of the mutated residue was chosen as the anchor atom. The backbone atoms (C-$\alpha$, N, and C) of the mutated residues were used as RBFE alignment atoms. The C-$\alpha$ atoms of the mutated residues were used as the origins of the respective ligand alignment frames. Quadratic flat-bottom positional restraints with a tolerance of $3.0$ \AA\ and a force constant of $25$ kcal/mol/\AA$^2$ were applied to the C-$\alpha$ atoms of the protein receptor and the bound peptide, except those of the mutated residue (residue 0) and the residue adjacent to it (residue 1) to allow for the wider range of conformational reorganization potentially caused by the mutation. Except for the RBFE alignment restraints, the unbound peptide was left unrestrained and free to explore the full ensemble of solution conformations.

The prepared protein-ligand and protein-peptide systems were energy-minimized, thermalized, and equilibrated at 300 K and 1 bar of constant pressure. This was followed by slow annealing to the $\lambda = 1/2$ alchemical intermediate for 250 ps. The resulting structure served as the initial configuration for the subsequent alchemical replica exchange simulations. We employed 22 replicas and the same alchemical schedule and softplus alchemical parameters as the CDK2 example of the AToM-OpenMM software.\cite{AToM-OpenMM-24} We used the softcore perturbation energy parameters $u_{\rm max} = 200$ kcal/mol, $u_c = 100$ kcal/mol, and $a = 1/16$ for all RBFE calculations. Asynchronous Hamiltonian replica exchange molecular dynamics conformational sampling\cite{gallicchio2015asynchronous} with the AToM-OpenMM software was carried out with a timestep of $2$ fs for $13.3$ ns/replica for the protein-ligand RBFEs and $40.0$ ns/replica for the protein-peptide RBFEs. Perturbation energy samples were collected every 40 ps. The UWHAM multi-state free energy estimator\cite{Tan2012,UHAM-in-R} was employed for free energy analysis after discarding the first third of the samples for equilibration. 

\section{Results}

We tested the novel Alchemical Transfer and coordinate Swapping (ATS) relative binding free energy (RBFE) protocol on the TYK2 community benchmark set of 24 ligand pairs binding the TYK2 tyrosine protein kinase\cite{liang2013lead-1,liang2013lead-2} assembled by Schr\"{o}dinger.\cite{wang2017accurate,ross2023maximal,Schrodinger-benchmarks} We further validated the ATS predictions against the experimental measurements (Table \ref{tab:tyk2-results}).  We further tested ATS on the protein-peptide RBFE benchmark studied by Panel et al.\cite{panel2018accurate,gallicchio2021comppeptsci}, which includes the binding of an 8-residue peptide and its single-point mutants to a PDZ domain. The much larger size of peptide ligands relative to small-molecule drug-like compounds provides a strict stress test of the hypothesis that the efficiency of the coordinate-swapping RBFE algorithm is independent of the ligands' size. In this case, a direct comparison of the ATS-RBFE and ATM-RBFE predictions is not feasible since the standard dual-topology ATM RBFE protocol does not apply to protein-peptide complexes. Instead, we compare to the single-topology estimates of Panel et al.\cite{panel2018accurate} and experimental binding affinities\cite{shepherd2011distinct,liu2013structure} (Table \ref{tab:pept-results} and Figure \ref{fig:pept-rbfe-diagram-1}).

All of the tests we conducted confirm the correctness of the novel ATS method and its applicability to estimating RBFEs for R-group transformations between large ligands and single-point mutants. 

\subsection{TYK-2 Protein-Ligand Benchmark}

The ATS-RBFE RBFE estimates for the TYK-2 benchmark closely agree with the standard ATM protocol (Table \ref{tab:tyk2-results}).  The root mean square deviation (RMSD) between the two sets is only $0.37$ kcal/mol, which is within statistical uncertainty, and the corresponding correlation coefficient is 91\%. Despite the distinct alchemical pathways, the close alignment between the ATS-RBFE and ATM-RBFE predictions strongly supports that they both faithfully reflect the true relative binding free energies of the molecular mechanics model of these systems. Moreover, the predictions correlate reasonably well with the experimental inhibition measurements (65 and 68\% correlation coefficients), confirming the relevance of our alchemical models for lead optimization in drug discovery.\cite{chen2024performance}    

\begin{table}
\caption{\label{tab:tyk2-results} The relative binding free energy estimates of the TYK2 pairs using the standard alchemical transfer (ATM-RBFE) and coordinate swapping (ATS-RBFE) workflows compared to the corresponding differences of experimental affinities.}
\begin{center}
\sisetup{separate-uncertainty}
\begin{tabular}{l S[table-format = 3.2(2)] S[table-format = 3.2(2)] c }
 Ligand Pair & \multicolumn{1}{c}{$\Delta\Delta G_b$$^{a,b}$} & \multicolumn{1}{c}{$\Delta \Delta G_b$$^{a,b}$} & $\Delta \Delta G_b$$^{a,c}$ \\ [0.5ex]
             & \multicolumn{1}{c}{(ATM-RBFE)}                 & \multicolumn{1}{c}{(ATS-RBFE)}                  & (Expt)                      \\ [0.5ex]
\hline
jmc23-ejm55    &  0.45(26) &  0.44(21) &  2.49 \\
ejm44-ejm55    & -2.55(27) & -2.25(23) & -1.79 \\
ejm49-ejm31    & -1.47(28) & -1.43(24) & -1.79 \\
ejm31-ejm46    & -0.82(26) & -0.80(20) & -1.77 \\
jmc28-jmc27    & -0.25(26) & -0.40(20) & -0.30 \\
ejm42-ejm48    &  0.82(27) &  0.01(22) &  0.78 \\
ejm31-ejm43    &  0.68(27) &  1.21(21) &  1.28 \\
ejm50-ejm42    & -0.91(26) & -0.78(20) & -0.80 \\
ejm42-ejm55    & -0.86(26) & -0.19(19) &  0.57 \\
jmc23-ejm46    & -0.38(26) & -0.40(20) &  0.39 \\
ejm31-ejm45    &  0.28(27) &  0.33(23) & -0.02 \\
ejm55-ejm54    & -0.32(27) & -0.44(23) & -1.32 \\
ejm45-ejm42    &  0.21(27) &  0.27(21) & -0.22 \\
ejm31-jmc28    &  0.19(27) & -0.27(22) & -1.44 \\
ejm31-ejm48    &  0.79(28) &  0.31(24) &  0.54 \\
ejm47-ejm31    & -0.76(27) & -0.19(22) &  0.16 \\
ejm47-ejm55    & -0.94(27) & -0.73(21) &  0.49 \\
ejm44-ejm42    & -2.64(27) & -2.21(22) & -2.36 \\
jmc23-jmc27    & -0.79(26) & -0.43(20) &  0.42 \\
ejm43-ejm55    & -1.75(26) & -1.48(21) & -0.95 \\
jmc23-jmc30    & -0.60(27) & -1.10(21) &  0.76 \\
jmc28-jmc30    & -0.74(27) & -1.29(22) &  0.04 \\
ejm42-ejm54    & -0.13(27) & -0.20(22) & -0.75 \\
ejm49-ejm50    & -1.08(28) & -0.84(25) & -1.23 \\
\hline
RMSD$^d$ vs.~ATM-RBFE &          & \multicolumn{1}{c}{$0.37$} &  \\
$R$$^e$  vs.~ATM-RBFE &          & \multicolumn{1}{c}{$0.91$} &  \\
\hline
RMSD$^d$ vs.~Expt & \multicolumn{1}{c}{$0.93$} & \multicolumn{1}{c}{$0.87$} & \\
$R$$^e$   vs.~Expt& \multicolumn{1}{c}{$0.65$} & \multicolumn{1}{c}{$0.68$} & \\
\hline
\end{tabular}
\begin{flushleft}\small
$^a$In kcal/mol. $^b$One standard deviation uncertainties in parenthesis.  $^c$From reference \citenum{ross2023maximal}. $^d$ Root mean square deviation in kcal/mol.  $^e$ Correlation coefficient.
\end{flushleft}
\end{center}
\end{table}

\subsection{TIAM-1 RBFE Protein-Peptide Benchmark}

\begin{figure}
    \centering
    \includegraphics[scale=1.0]{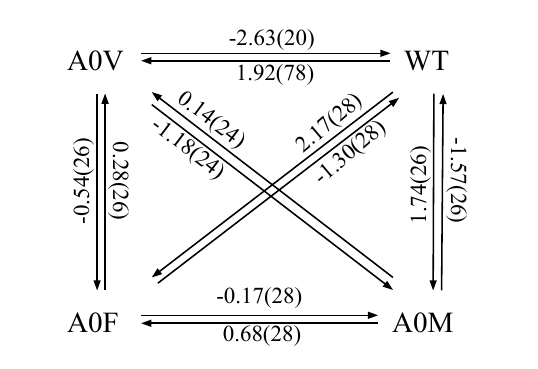}
    \caption{\label{fig:pept-rbfe-diagram-1} Diagram of the ATS-RBFE estimates from Table \ref{tab:pept-results}. The values are in kcal/mol, and the statistical uncertainties are indicated in parenthesis.}
\end{figure}

We computed the relative binding free energies between all pairs and in both alchemical directions of the wild-type Syndecan-1-derived peptide and three single-point mutant at the first position (see Molecular Systems) investigated by Panel et al.\cite{panel2018accurate} 
 The resulting values, listed in Table \ref{tab:pept-results} and illustrated in Figure \ref{fig:pept-rbfe-diagram-1}, are in qualitative agreement with the values reported by Panel et al.\cite{panel2018accurate} and the experimental affinities,\cite{panel2017simple} where they are available. Even though the force field model we employed appears to generally overestimate the loss of affinity of the mutants, the calculations confirm the higher affinity of the wild-type peptide over the three mutants at the first position. Moreover, the predicted relative ranking of the A0M and A0F mutants is reversed relative to the experiments.
  
  In addition to the qualitative alignment with the experiments, the ATS estimates have a high degree of self-consistency, indicating that the estimates are converged and reflective of the model's accuracy. The average pairwise hysteresis errors calculated from the sum of the relative free energy estimates of each pair in the two directions ($0.59$ kcal/mol) is within statistical uncertainty--but is as high as $1.0$ kcal/mol for the A0M/A0V pair. The average of the absolute values of the cycle closure errors over the cycles of length 3 (8 cycles) and 4 (6 cycles) are small ($0.52$ and $0.23$ kcal/mol, respectively). The reduction of cycle closure errors for larger cycles is indicative of random statistical RBFE uncertainties that tend to cancel rather than systematic errors in the calculation.

\begin{table}
\caption{\label{tab:pept-results}  The alchemical transfer coordinate swapping relative binding free energy estimates of the complexes between the TIAM-1 PDZ domain and the wild-type Syndecan-1 peptide and its mutants compared to the literature values and the corresponding differences of the available experimental affinities.}
\begin{center}
\sisetup{separate-uncertainty}
\begin{tabular}{l S[table-format = 3.2(2)] S[table-format = 3.2(2)] c }
 Peptide Pair & \multicolumn{1}{c}{$\Delta\Delta G_b$$^{a,b,c}$} & \multicolumn{1}{c}{$\Delta \Delta G_b$$^{a,b,d}$} & $\Delta \Delta G_b$$^{a,d}$ \\ [0.5ex]
             & \multicolumn{1}{c}{(ATS-RBFE)}                  & \multicolumn{1}{c}{(Panel et al)}               & (Expt)$^e$                    \\ [0.5ex]
\hline
WT-A0M    &  1.74(26)  &  1.80(50)  &    1.56 \\
A0M-WT    & -1.57(26)  &            &   -1.56 \\
WT-A0V    &  1.92(78)  &  1.90(10)  &         \\
A0V-WT    & -2.63(20)  &            &          \\
WT-A0F    &  2.17(28)  &  0.50(100) &    0.43 \\
A0F-WT    & -1.30(28)  &            &   -0.43 \\
A0M-A0F   &  0.68(28)  & -1.60(10)  &   -1.13 \\
A0F-A0M   & -0.17(28)  &            &    1.13 \\
A0V-A0F   & -0.54(26)  & -3.20(10)  &    -   \\
A0F-A0V   &  0.28(26)  &            &    -   \\
A0M-A0V   &  0.14(24)  &            &    -   \\
A0V-A0M   & -1.18(24)  &            &    -  \\

\end{tabular}
\begin{flushleft}\small
$^a$In kcal/mol. $^b$One standard deviation uncertainties in parenthesis. $^c$ This work. $^d$ From reference \citenum{panel2018accurate}. $^e$ From reference \citenum{panel2017simple}.
\end{flushleft}
\end{center}
\end{table}

\section{Discussion}

In this work, we developed and tested a dual-topology algorithm based on the Alchemical Transfer Method (ATM) called Alchemical Transfer with Coordinate Swapping (ATS), valid for any molecular transformation involving peripheral R-groups where a single-topology method applies.\cite{Mey2020Best} In contrast to the ATM-RBFE protocol that translates the unbound ligand into the binding site and the bound ligand to the solution as a whole,\cite{azimi2022relative} ATS translates only the two R-groups that differ between the two ligands. The structure of covalent interactions of the two ligands is automatically preserved by swapping the coordinates of the atoms of the ligands' common regions unaffected by the alchemical transformation. 

The method retains the simplicity and versatility of a dual-topology approach\cite{Michel2007,riniker2011comparison} while affording the same convergence rate of single-topology that scales as the size of the transformation rather than the size of the ligands.\cite{pearlman1994comparison} We validated the method's applicability and correctness by comparing its predictions to those of the standard ATM-RBFE protocol on the congeneric TYK2 RBFE benchmark,\cite{ross2023maximal} for which ATS yielded RBFE estimates equivalent to ATM but with lower statistical uncertainty with the same computational cost.  We also illustrated using ATS to estimate the RBFEs between mutants of protein-peptide complexes where ATM is not applicable.\cite{panel2018accurate} Here, ATS yielded converged RBFE estimates in line with experiments and single-topology calculations reported in the literature. These results confirm the correctness of the ATS implementation and its usefulness in improving the efficiency of the alchemical transfer technology and helping extend it to macromolecular complexes.

While it solves some key shortcomings of the dual-topology protocol, ATS does not directly address the widespread challenge in alchemical calculations of obtaining equilibrated conformational ensembles at the physical end states and along the alchemical pathway.\cite{procacci2019solvation} Conformational sampling of protein-peptide complexes is particularly difficult due to their flexibility and extensive conformational reorganization.\cite{Gallicchio2021binding}
Furthermore, the ATS extension does not address alchemical transfer's lower time-step computational performance relative to double-decoupling single-topology implementations. ATS and ATM require larger simulation boxes to accommodate the unbound ligand and, being based on an energy interpolation scheme, evaluate the potential energy function twice at each step. However, a definitive assessment of the relative effective computational performance of the two approaches is not obvious since alchemical transfer requires fewer independent calculations and employs a more direct alchemical path that could yield faster convergence of free energy estimates. In addition, like ATM, ATS supports any energy function, including advanced many-body potentials such as neural network potentials, quantum-mechanical models, and polarizable force fields, which have started to be employed in drug-discovery applications.\cite{wang-pearlman2021quantum,sabanes2024enhancing,silvestri2024targeting,ansari2024targeting-chemarxiv} An exciting prospect is the application of ATS in conjunction with our recently developed receptor-hopping protocol to study the binding specificity of macromolecular ligands.\cite{azimi2024binding}

\section{Conclusion}

We presented the Alchemical Transfer with Coordinate Swapping (ATS) method to enable the calculation of the relative binding free energies between large ligands that differ by a small R-group with the Alchemical Transfer Method (ATM) framework. The new method works by transferring only the portion of the molecules that differ between the bound and unbound ligands. The common region of the two ligands, which can be arbitrarily large, is unchanged and does not contribute to the magnitude and statistical fluctuations of the perturbation energy. However, internally, the coordinates of the atoms of the common regions are swapped to maintain the integrity of the covalent bonding data structures of the molecular dynamics engine. We successfully validated the method on protein-ligand and protein-peptide RBFE benchmarks. This advance paves the road for applying the relative binding free energy Alchemical Transfer Method protocol to study the effect of protein and nucleic acid mutations on the stability of macromolecular complexes. Future work will also investigate the applicability of ATS in conjunction with receptor-hopping protocols to study the binding specificity of macromolecular ligands.

\section*{Appendix}

\subsection{Proof of the coordinate swapping relative binding free energy formula}

The statistical mechanics expression for the bimolecular binding constant between a receptor $R$ and a ligand $A$ is\cite{Gilson:Given:Bush:McCammon:97,Gallicchio2011adv,Gallicchio2021binding}
\begin{equation}
K_{b}(A) =\frac{C^{\circ}}{8\pi^{2}}\frac{z_{RA}}{z_{R}z_{A}},
\label{eq:Kb_statmech}
\end{equation}
where $z_{RA}$, $z_{R}$, and $z_{A}$, the internal partition functions of the complex, receptor, and ligand, respectively, are defined as
\begin{equation}
z_{A} = \int dx_A  e^{-\beta \Psi_{A}(x_{A})}
    \label{eq:part-func-zR}
\end{equation}
and similarly for R,
\begin{equation}
z_{RA} = \int dx_R dx_A d\zeta_A I(\zeta_A) e^{-\beta \Psi_{RA}(x_{R}, x_{A}, \zeta_A)}
    \label{eq:part-func-zRL}
\end{equation}
where $x_R$ and $x_A$ are the internal coordinates of the receptor and ligand respectively. $\zeta_A = (\bm{c}_A, \bm{\omega}_A)$ denotes, collectively, the three position and three orientation coordinates of ligand $A$ relative to the reference frame of the receptor, $\Psi_A$ and $\Psi_{RA}$ denote the effective potential energy functions in the solvent potential of mean force representation\cite{Roux:Simonson:99,Gallicchio2021binding} of the ligand in the solvent and of the ligand bound to the solvated receptor, respectively. The function $I(\zeta_A)$ is defined later. The function $\Psi$ is the potential energy function of the system in the solvent potential of mean force representation, obtained by pre-averaging over the solvent degrees of freedom. The potential of mean force representation does not introduce approximations; it is used here and in the following derivations as for notational convenience to avoid explicitly listing the coordinates of the solvent.

In Eq.\ (\ref{eq:part-func-zRL}), the function $I(\zeta_A)$ in Eq.\ (\ref{eq:part-func-zRL}) is an indicator function that is set to $1$ if the  position and orientation of the
ligand are such that receptor and ligand are considered bound, and zero otherwise.\cite{Gilson:Given:Bush:McCammon:97,Boresch:Karplus:2003,Gallicchio2011adv} The volume of configurational space encompassed by the indicator function is denoted here by $V_{\rm site} \Omega_{\rm site}$:
\begin{equation}
\int d\zeta I(\zeta) = V_{\rm site} \Omega_{\rm site}
\label{eq:vsite}
\end{equation}
For later use, we consider also an indicator function $I^\ast(\zeta)$ identical to the one that defines the complex, but centered at a point $\bm{d}$ in the solvent bulk relative to the reference frame of the receptor. 

Consider now the ratio, $K_b({\rm B})/K_b({\rm A})$, of the equilibrium binding constants of two ligands $A$ and $B$ to the same receptor $R$ in the same binding mode described by the indicator function $I(\zeta)$. To simplify the notation, in the following we will also assume that  $I(\zeta)$ is independent of the orientation coordinates. That is we assume that $I(\zeta) = I(\bm{c})$, where $\bm{c}$ is a chosen centroid of the ligand.  Under this assumption $\Omega_{\rm site} = 8 \pi^2$ which cancels the same factor in the denominator of Eq.\ (\ref{eq:Kb_statmech}). The extension of this derivation to orientation-dependent binding mode definitions is straightforward and does not affect the end result.

From Eq.\ (\ref{eq:Kb_statmech}) and canceling the common factor $z_{\rm R}$, we have
\begin{equation}
    \frac{K_b({\rm B})}{K_b({\rm A})} =
    \frac{z_{{\rm RB}} z_{\rm A}}{z_{{\rm RA}} z_{\rm B}} \, ,
    \label{eq:Kbratio1}
 \end{equation}
 where $z_{\rm RB}$ is the intramolecular configurational partition function of the complex between R and B, $z_{\rm B}$ is the intramolecular partition function of ligand B, and similarly for $z_{\rm RA}$ and $z_{\rm A}$. To express Eq.\ (\ref{eq:Kbratio1}) as an ensemble average the products $z_{{\rm RB}} z_{\rm A}$ and  $z_{{\rm RA}} z_{\rm B}$ are expressed as partition function integrals of systems containing the receptor and the two ligands such that one of the ligands is in the binding site and the other is centered on a position of the bulk at distance $\bm{d}$ relative to the receptor coordinate frame. Specifically, we multiply the numerator of Eq.\ (\ref{eq:Kbratio1}) by Eq.\ (\ref{eq:vsite}) for A and the denominator by the corresponding integral for B. The result is
  \begin{equation}
    \frac{K_b(B)}{K_b(A)} =
    \frac{
    \int dx_R dx_A dx_B d\bm{c}_A d\bm{\omega}_A d\bm{c}_B d\bm{\omega}_B I^\ast(\bm{c}_A) I(\bm{c}_B)  e^{-\beta \Psi(x_{R}, x_{A}, x_B, \bm{c}_A, \bm{\omega}_A, \bm{c}_B, \bm{\omega}_B)}
    }{
    \int dx_R dx_A dx_B d\bm{c}_A d\bm{\omega}_A d\bm{c}_B d\bm{\omega}_B I(\bm{c}_A) I^\ast(\bm{c}_B) e^{-\beta \Psi(x_{R}, x_{A}, x_B, \bm{c}_A, \bm{\omega}_A, \bm{c}_B, \bm{\omega}_B)}
    }
    \label{eq:Kb-statmech-rel-1}
\end{equation}
Next, we express the integration variables in terms of the Cartesian coordinates of the atoms $\bm{r}_A$ calculated from the internal coordinates $x_A$, the centroid coordinates $\bm{c}_A$, and the orientational coordinates $\omega_A$, and similarly for $B$. The centroids becomes a function of the ligands' Cartesian coordinates, which are constrained to be within the region spanned by each indicator function. In the denominator for example, the center of mass of $B$ is constrained by $I^\ast()$, which is located in the solvent bulk. Hence, the coordinates of $B$ are constrained to be near this location.
    \begin{equation}
    \frac{K_b(B)}{K_b(A)} =
    \frac{
    \int dx_R d\bm{r}_A d\bm{r}_B I^\ast(\bm{c}_A) I(\bm{c}_B) e^{-\beta \Psi(x_{R}, \bm{r}_{A}, \bm{r}_B)}
    }{
    \int dx_R d\bm{r}_A d\bm{r}_B I(\bm{c}_A) I^\ast(\bm{c}_B) e^{-\beta \Psi(x_{R}, \bm{r}_{A}, \bm{r}_B)}
    }
    \label{eq:Kb-statmech-rel-1-cartesian}
\end{equation}
Next, we divide the atoms of the ligands into common subsets $A'$ and $B'$ and variable subsets $A''$ and $B''$. That is, for example, $\bm{r}_A = (\bm{r}_{A'}, \bm{r}_{A''})$, where  $\bm{r}_{A'}$ are the coordinates of the atoms in the common subset of $A$ and $\bm{r}_{A''}$ those of the atoms in the variable subset, and similarly for ligand $B$. The common subsets for $A$ and $B$ have the same dimension. Additionally, a one-to-one and invertible (bijective) mapping relationship is established between the atoms in the common subset of $A$ and the corresponding atoms of $B$. The atoms in the variable subsets are assumed to belong to two corresponding sidechain of the molecule each attached to an anchoring atom, $a$ of $A$ and $b$ for $B$, belonging to the common subset and mapped into each other.

Finally, to express Eq.\ (\ref{eq:Kb-statmech-rel-1}) as an ensemble average, we perform the following change of variables in the integral at the numerator of Eq.\ (\ref{eq:Kb-statmech-rel-1-cartesian}): (i) displacement for the atoms in the variable subsets by the vector distance, $\bm{d}_{BA} = \bm{r}_b - \bm{r}_a \simeq \bm{d}$ between the two anchoring atoms, and 
\begin{eqnarray}
  \bm{r}_{A''} & \rightarrow & \bm{r}_{A''} + \bm{d}_{BA}    \nonumber \\
  \bm{r}_{B''} & \rightarrow & \bm{r}_{B''} - \bm{d}_{BA} \label{eq:displ-transformation}
\end{eqnarray}
(ii) swapping of the coordinates of the corresponding atoms in the two common subsets
\begin{eqnarray}
  \bm{r}_{A'} & \rightarrow & \bm{r}_{B'}     \nonumber \\
  \bm{r}_{B'} & \rightarrow & \bm{r}_{A'} \label{eq:swapping-transformation}
\end{eqnarray}
which in overall move the coordinates of $B$ from the solvent to the binding site and the coordinates of $A$ from the binding site to the solvent, but maintaining the internal coordinates of the common regions unchanged.

As a result of the transformations above, the centroid of $A$, forced to be within the solvent region by the indicator function $I^\ast()$, is moved to the solvent region corresponding to the indicator function $I()$. Assuming that $I()$ is large enough so that a centroid in one region always lands in the allowed region of the other as a result of the transformations and viceversa, the value of the product $I^\ast(\bm{c}_A) I(\bm{c}_B)$ is not affected by the transformation. Furthermore, because after the transformation approximately $I^\ast(\bm{c}_A) \rightarrow I^\ast(\bm{c}_A + \bm{d} ) = I(\bm{c}_A)$, and similarly for $B$, this term becomes $I(\bm{c}_A) I^\ast(\bm{c}_B)$, matching the same term in the integral at the denominator of Eq.\ (\ref{eq:Kb-statmech-rel-1-cartesian}).

The absolute value of the determinant $|J|$ of the Jacobian of the variable transformations (\ref{eq:displ-transformation}) and (\ref{eq:swapping-transformation}) is 1. This can shown by considering that the gradient of the coordinates resulting from the swapping transformation (\ref{eq:swapping-transformation}) relative to the original coordinates is a vector of zeros except for the term corresponding to the mapped atom, where it is 1. Similarly, the gradient for the displacement transformation (\ref{eq:displ-transformation}) is 1 on the diagonal and $1$ or $-1$ in correspondence with the anchoring atom. Finally, the transformation does not affect the coordinates of the receptor atoms, which are represented by an identity matrix block of the Jacobian. 

For example, consider a system composed of a receptor with two atoms, a ligand $A$ with four atoms, two in the common subset and two in the variable subset, and a ligand $B$ made of three atoms with two in the common subset and one in the variable subset. Without loss of generality, we index the atoms of the system so that the receptor is listed first followed by the ligand atoms so that the two corresponding anchoring atoms have the last two indexes (8 and 9). Ligand $A$ has atoms 3 and 8 in the common subset and atoms 4 and 5 in the variable subset and ligand $B$ has atoms 6 and 9 in the common subset and atom 7 in the variable subset. Atom 3 is mapped to atom 6 and atom 8 is mapped to atom 9. This arrangement results in the following coordinate transformation
\begin{eqnarray*}
  \bm{r}'_1 & = & \bm{r}_1 \\
  \bm{r}'_2 & = & \bm{r}_2 \\
  \bm{r}'_3 & = & \bm{r}_6 \\
  \bm{r}'_4 & = & \bm{r}_4 + \bm{r}_9 - \bm{r}_8 \\
  \bm{r}'_5 & = & \bm{r}_5 + \bm{r}_9 - \bm{r}_8 \\
  \bm{r}'_6 & = & \bm{r}_3 \\
  \bm{r}'_7 & = & \bm{r}_7 - \bm{r}_9 + \bm{r}_8 \\
  \bm{r}'_8 & = & \bm{r}_9 \\
  \bm{r}'_9 & = & \bm{r}_8
\end{eqnarray*}
where the primed symbols are the coordinates after the transformation. The Jacobian matrix $\partial \bm{r}'_i/\partial \bm{r}_j$ of the transformation above is (we show only the components for one of the coordinate axis since the coordinate transformation does not include mixed terms)
\begin{equation*}
\begin{pmatrix}
  1 & 0 & 0 & 0 & 0 & 0 & 0 & 0 & 0 \\
  0 & 1 & 0 & 0 & 0 & 0 & 0 & 0 & 0 \\
  0 & 0 & 0 & 0 & 0 & 1 & 0 & 0 & 0 \\
  0 & 0 & 0 & 1 & 0 & 0 & 0 &-1 & 1 \\
  0 & 0 & 0 & 0 & 1 & 0 & 0 &-1 & 1 \\
  0 & 0 & 1 & 0 & 0 & 0 & 0 & 0 & 0 \\
  0 & 0 & 0 & 0 & 0 & 0 & 1 & 1 &-1\\
  0 & 0 & 0 & 0 & 0 & 0 & 0 & 0 & 1 \\
  0 & 0 & 0 & 0 & 0 & 0 & 0 & 1 & 0
\end{pmatrix}
\end{equation*}
Because the swapping transformation is one-to-one, it is always possible to swap the columns of the Jacobian matrix so that each column has a 1 in the diagonal corresponding to either the original atom (displacement) or the mapped atom (swapping). This involves swapping columns 8 and 9 that correspond to the anchoring atoms. However, because the non-diagonal terms of these columns are above the diagonal ones (because the anchoring atoms affect the transformations of only the atoms with lower index), these terms remain in the upper triangular portion of the Jacobian matrix. The result of swapping the columns, which does not change the absolute value of the determinant, yields
\begin{equation*}
\begin{pmatrix}
  1 & 0 & 0 & 0 & 0 & 0 & 0 & 0 & 0 \\
  0 & 1 & 0 & 0 & 0 & 0 & 0 & 0 & 0 \\
  0 & 0 & 1 & 0 & 0 & 0 & 0 & 0 & 0 \\
  0 & 0 & 0 & 1 & 0 & 0 & 0 & 1 &-1 \\
  0 & 0 & 0 & 0 & 1 & 0 & 0 & 1 &-1 \\
  0 & 0 & 0 & 0 & 0 & 1 & 0 & 0 & 0 \\
  0 & 0 & 0 & 0 & 0 & 0 & 1 &-1 & 1 \\
  0 & 0 & 0 & 0 & 0 & 0 & 0 & 1 & 0 \\
  0 & 0 & 0 & 0 & 0 & 0 & 0 & 0 & 1
\end{pmatrix}
\end{equation*}
The matrix above is an upper triangular matrix in which the diagonal blocks are identity matrices with unitary determinant and therefore its determinant is 1.\cite{Taboga2021}

With the preparation above, Eq.\ (\ref{eq:Kb-statmech-rel-1-cartesian}) is rewritten as:
    \begin{equation}
    \frac{K_b(B)}{K_b(A)} =
    \frac{
    \int dx_R d\bm{r}_{A'} d\bm{r}_{A''} d\bm{r}_{B'} d\bm{r}_{B''}  I(\bm{c}_A) I^\ast(\bm{c}_B) e^{-\beta \Psi(x_{R}, \bm{r}_{B'},  \bm{r}_{A''}+\bm{d_{BA}},  \bm{r}_{A'}, \bm{r}_{B''}-\bm{d}_{BA})}
    }{
    \int dx_R d\bm{r}_{A'} d\bm{r}_{A''} d\bm{r}_{B'} d\bm{r}_{B''}  I(\bm{c}_A) I^\ast(\bm{c}_B) e^{-\beta \Psi(x_{R}, \bm{r}_{A'},  \bm{r}_{A''},  \bm{r}_{B'}, \bm{r}_{B''})}
    }
    \label{eq:Kb-statmech-rel-trans-1}
\end{equation}

Finally, Eq.\ (\ref{eq:Kb-statmech-rel-trans-1}) is recovered by multiplying and dividing the integrand in the numerator by the Boltzmann factor in the denominator, . 

\subsection{Gradients of the Alchemical Potential Energy Function}

The gradient of the potential in Eq.\ (\ref{eq:pot-func}) with respect to the coordinate $\bm{r}_k$ of an atom is
\begin{equation}
    \frac{\partial U_\lambda}{\partial \bm{r}_k} = \frac{\partial U_{RA+B}}{\partial \bm{r}_k} + W'(u) \frac{\partial u}{\partial \bm{r}_k}
    \label{eq:dpot-1}
\end{equation}
where $W'(u)$ is the derivative of the alchemical perturbation function (\ref{eq:softplus-function}), and the unperturbed potential energy function $U_{RA+B}$, the perturbed potential energy function $U_{RB+A}$, and the perturbation energy $u$ are defined by Eqs.\ (\ref{eq:pert-func-rel})--(\ref{eq:pot-Bbound}). Eq.\ (\ref{eq:dpot-1}) is then expressed as a linear combination of the gradients of the perturbed and unperturbed potentials:
\begin{equation}
    \frac{\partial U_\lambda}{\partial \bm{r}_k} = [1 - W'(u)] \frac{\partial U_{RA+B}}{\partial \bm{r}_k} + W'(u) \frac{\partial U_{RB+A}}{\partial \bm{r}_k}
    \label{eq:dpot-2}
\end{equation}
The gradient of the unperturbed potential energy function is collected by the molecular dynamics engine as usual. The perturbed potential energy function is the unperturbed one with the transformed coordinates (displaced and swapped),
\begin{equation}
U_{RB+A}(\bm{r}) = U_{RA+B}[\bm{r}'(\bm{r})]
\end{equation}
hence, its gradients can be found by the chain rule
\begin{equation}
\frac{\partial U_{RB+A}}{\partial \bm{r}_k} = \sum_j \frac{\partial U_{RA+B}(\bm{r}')}{\partial \bm{r}'_j}  \frac{\partial \bm{r}'_j}{\partial \bm{r}_k}
\label{eq:chain-rule}
\end{equation}
The first term in the sum above is the gradient of the system's potential energy function computed after the variable transformation, also available from the molecular dynamics engine.

If $k$ refers to one of the atoms in the common regions of $A'$ or $B'$ other than the anchoring atom, only the coordinates of its mapped atom, $k'$, in the transformed system depend on it. Hence, only the $j=k'$ term in the sum (\ref{eq:chain-rule}) is not zero. Furthermore because $\bm{r}'_{k'} = \bm{r}_{k}$, ${\partial \bm{r}'_{k'}}/{\partial \bm{r}_k}$ is the identity matrix. So we have:
\begin{equation}
\frac{\partial U_{RB+A}}{\partial \bm{r}_k} = \frac{\partial U_{RA+B}(\bm{r}')}{\partial \bm{r}'_{k'}} \quad k \in A', B'
\end{equation}
which states that the gradient of $U_{RB+A}$ with respect to atom $k$ is the gradient of the potential energy of the transformed system with respect to its mapped atom. 

Let us now consider an atom $k$ in the variable region $A''$ of ligand $A$. In the transformed system, atom $k$ is translated by the displacement vector of the two anchoring atoms, $\bm{r}'_k = \bm{r}_k + \bm{r}_b - \bm{r}_a$, and because neither the positions of the anchoring atoms nor the coordinates of any other atom depend on it:
\begin{equation}
\frac{\partial U_{RB+A}}{\partial \bm{r}_k} = \frac{\partial U_{RA+B}(\bm{r}')}{\partial \bm{r}'_{k}} \quad k \in A'', B''
\end{equation}
where we observed that the same argument applies to one of the atoms in the variable region of $B$.

Finally, consider the anchoring atom of $A$, i.e.\ $k=a$ in Eq.~(\ref{eq:chain-rule}). All of the transformed coordinates of the atoms in the variable regions of $A$ and $B$ depend on it through the displacement vector, $\bm{d}_{BA }=\bm{r}_b - \bm{r}_a$ for atoms in $A''$ and the opposite displacement vector for atoms in $B''$.  Moreover, $\bm{r}'_b = \bm{r}_a$ because $a$ in the common region of $A$. It follows that the terms $j=b$, and those for $j \in A''$ (with a negative sign) and for $j \in B''$ (with a positive sign) are not zero in Eq.~(\ref{eq:chain-rule}):
\begin{equation}
\frac{\partial U_{RB+A}}{\partial \bm{r}_a} = \frac{\partial U_{RA+B}(\bm{r}')}{\partial \bm{r}'_b} +  \sum_{j \in B''} \frac{\partial U_{RA+B}(\bm{r}')}{\partial \bm{r}'_j} - \sum_{j \in A''} \frac{\partial U_{RA+B}(\bm{r}')}{\partial \bm{r}'_j}
\label{eq:chain-rule-a}
\end{equation}
The same applies to the anchoring atom $b$, but with the sign reversed:
\begin{equation}
\frac{\partial U_{RB+A}}{\partial \bm{r}_b} = \frac{\partial U_{RA+B}(\bm{r}')}{\partial \bm{r}'_a} -  \sum_{j \in B''} \frac{\partial U_{RA+B}(\bm{r}')}{\partial \bm{r}'_j} + \sum_{j \in A''} \frac{\partial U_{RA+B}(\bm{r}')}{\partial \bm{r}'_j}
\label{eq:chain-rule-b}
\end{equation}
In our implementation, the sums of the gradients in Eqs.~(\ref{eq:chain-rule-a}) and (\ref{eq:chain-rule-b}) above are collected when scanning the gradients of the atoms. If an atom $j$ is part of one of the variable regions, its gradient is added to those of the anchoring atoms with the correct sign depending on whether the atom belongs to $A''$ or $B''$ and whether the gradient of anchoring atom $a$ or $b$ is being updated.

\bibliography{main,add2main}

\end{document}